\documentstyle[12pt,epsf]{article}

\makeatletter
\@addtoreset{equation}{section}
\makeatother

\newcommand{\ra}{\rightarrow}

\newcommand{\cF}{{\mathcal F}}
\newcommand{\cH}{{\mathcal H}}
\newcommand{\del}{\partial}
\newcommand{\half}{\frac{1}{2}}

\newcommand{\beq}{\begin{eqnarray}}
\newcommand{\eeq}{\end{eqnarray}}
\newcommand{\beqa}{\begin{eqnarray}}
\newcommand{\eeqa}{\end{eqnarray}}
\newcommand{\pa}{\partial}

\begin{document}

\begin{titlepage}
\vspace*{-2.5cm}
\null
\begin{flushright}
hep-th/0205085 \\
CITUSC/02-015\\
ITFA-2002-14\\
May, 2002
\end{flushright}
\vspace{0.5cm}
\begin{center}
{\Large \bf
Tachyon Matter in Boundary String Field Theory\\
\par}
\lineskip .75em
\vskip1.5cm
\normalsize

{\large
Shigeki Sugimoto\footnote{
E-mail:\ \ sugimoto@citusc.usc.edu}  and
Seiji Terashima\footnote{
E-mail:\ \ sterashi@science.uva.nl} }
\vskip 2.5em
{
${}^1$
 \it
CIT/USC Center for Theoretical Physics,
University of Southern California,\\
 Los Angeles CA90089-2535, USA \\}
\vskip 1em
{
${}^2$
 \it  Institute for Theoretical Physics,
University of Amsterdam \\
Valckenierstraat 65,
1018 XE, Amsterdam, The Netherlands}
\vskip 3em

{\it Dedicated to the memory of Sung-Kil Yang}
\vskip 2em

{\bf Abstract}
\end{center}

We analyse the classical decay process of
unstable D-branes in superstring theory using
the boundary string field theory (BSFT) action.
We show that the solutions of the equations of motion
for the tachyon field
asymptotically approach to $T=x^0$ and
the pressure rapidly falls off at late time
producing the tachyon matter
irrespective of the initial condition.
We also consider the cosmological evolution
driven by the rolling tachyon
using the BSFT action as an effective action.

\end{titlepage}

\baselineskip=0.7cm

\section{Introduction}

Recently, classical time dependent solutions
in string theory describing
the decay process of an unstable D-brane
was constructed \cite{Se1,Se2}.
The decay of an unstable D-brane is
caused by the tachyon field on the brane
which is rolling down toward
the minimum of the potential.
In particular, it was shown in \cite{Se2} that
the rolling of tachyon makes the unstable D-brane
toward the tachyon matter which
is pressureless gas with nonzero energy density.
These solutions are described by boundary states
of the unstable D-brane including boundary perturbation
associated with the rolling tachyon.
In \cite{Se3,Se2}, it was also shown that
the effective action proposed by \cite{Ga,Be,Kl}
with a specific tachyon potential
reproduces the following two properties of the tachyon matter:
1) absence of plane wave solutions, and
2) exponential fall off of the pressure.
However, since this action with the potential
has not been derived from any off-shell string theory,
the validity of the analysis based on this
action is not clear.

On the other hand,
the action of boundary string field theory (BSFT) for
bosonic string theory \cite{Wi}
successfully describes some dynamics of the tachyon
\cite{GeSh,KuMaMo1}.
In the superstring case,
the BSFT actions were obtained for
non-BPS D-branes \cite{KuMaMo2,Ma,NiPr} and
D-brane- anti D-brane pairs \cite{KrLa,TaTeUe}.
These actions exactly describe lower dimensional D-branes
as solitons and
are expected to satisfy the first property above
because of their forms of tachyon potentials.
Moreover, we know the coupling of massless closed string modes,
including the RR-fields \cite{KrLa,TaTeUe},
to the fields in the unstable D-brane in BSFT.
Therefore it is both reasonable and interesting to consider
the rolling of the tachyon in BSFT.

In this paper, we analyse the classical decay
process of the unstable D-brane using the BSFT action.
More precisely, we will use the usual
BSFT action which is obtained by neglecting
$\del \del T$ and higher derivatives.
This action is exact as long as we use the
linear tachyon profile $T=v+\sum u_\mu x^\mu$.
\footnote{In this paper, we only consider the superstring.}

At first sight, one might think that
the rolling tachyon will hit the singularity within finite
time, since the geodesic distance between the points
$T=0$ and $T=\infty$ in the space of tachyon field $T$ is
finite \cite{KuMaMo1}. However, as we will show
in the following section,
the time dependent solutions
of the action approach $T\sim \pm x^0$ asymptotically
and keep rolling forever.
The solution with the asymptotic behavior such that
$T\sim x^0$ as $x^0\ra \pm\infty$
can be thought of as the space-like brane considered
in \cite{GuSt}.

We will also show that
the pressure of the system
will rapidly converge to zero while the energy density
is conserved. Hence the system will become the
tachyon matter which is consistent with the Sen's result.

Unfortunately,
since our solution is slightly away from
the linear profile, where
the BSFT action is reliable,
some of our results may receive corrections.
We will discuss this issue in section \ref{Conc}.

One of the interesting subject
for the rolling tachyon of unstable D-brane systems
is the relevance in cosmology
\cite{Gi}--\cite{ShWa}.
We will also consider
the system coupled with gravity,
and examine cosmological evolution
using the BSFT action.

The rest of the paper is organized as follows.
In section \ref{Tmatter}, we analyse the
BSFT action for a non-BPS D-brane,
and obtain the asymptotic solution of
the rolling tachyon. We show that the system
will become the tachyon matter in the large time limit.
In section \ref{Cosmo},
we couple the system to gravity
and examine the cosmological evolution.
Section \ref{Conc}
is devoted to conclusion and discussion.

\section{Tachyon matter in BSFT}
\label{Tmatter}

In this paper we consider the non-BPS D$p$-brane
in type II string theory.
The effective theory of non-BPS D$p$-brane is described by
the BSFT action \cite{KuMaMo2}
\beq
S=-\tilde{T}_p \int d^{p+1} x\, e^{-\frac{T^2}{4}}
{\mathcal F}\left( \frac{\alpha'}{2}\eta_{\mu \nu} \pa_\mu T \pa_\nu T \right),
\label{bsfta}
\eeq
where
\beq
{\mathcal F}(z)=z \frac{4^z}{2} \frac{\Gamma(z)^2}{\Gamma(2z)}
=\frac{\sqrt{\pi}\Gamma(z+1)}{\Gamma(z+1/2)}
\eeq
and $\tilde{T}_p$ is the tension of the non-BPS D$p$-brane.
Here we only consider the tachyon field and set the other
fields to zero for simplicity.
This action is considered to be
exact if we set $T=v+\sum u_\mu x^\mu$,
where $v$ and $u_\mu$ are constants.
Using the expansion
\beq
{\mathcal F}(z)=1+2 \log2\, z +{\cal O}(z^2),
\label{asF}
\eeq
we can read the potential and kinetic term for the tachyon
\beq
S=-\tilde{T}_p \int d^{p+1}x\, e^{-\frac{T^2}{4}} \left(
1+ \log2 \, \alpha'\pa_\mu T \pa^\mu T
+{\cal O}((\pa_\mu T \pa^\mu T)^2) \right).
\label{kinpo}
\eeq

\begin{figure}[htb]
\unitlength=.4mm
\begin{center}
\parbox{5cm}{
\begin{picture}(150,100)(0,0)
\epsfxsize=5cm
\put(0,0){\epsfbox{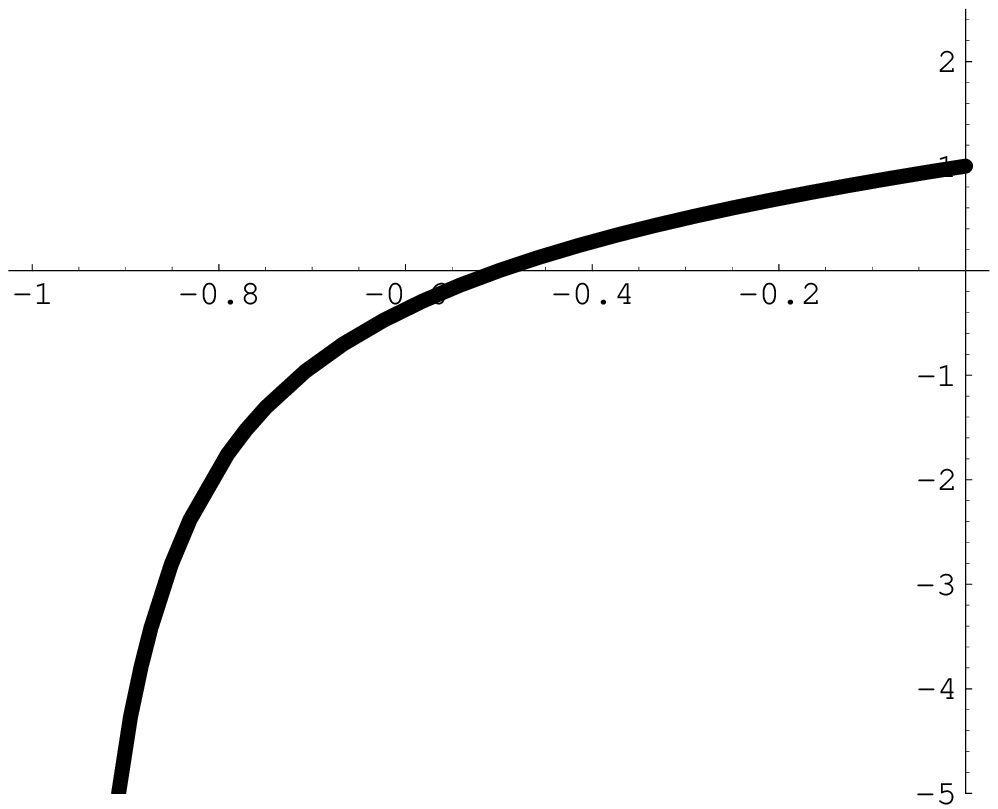}}
\put(137,100){\makebox(0,0){$\cF(z)$}}
\put(130,68){\makebox(0,0){$z$}}
\end{picture}
\caption{}
\label{F.eps}
}\hspace{10ex}
\parbox{5cm}{
\begin{picture}(150,100)(0,0)
\epsfxsize=5cm
\put(0,0){\epsfbox{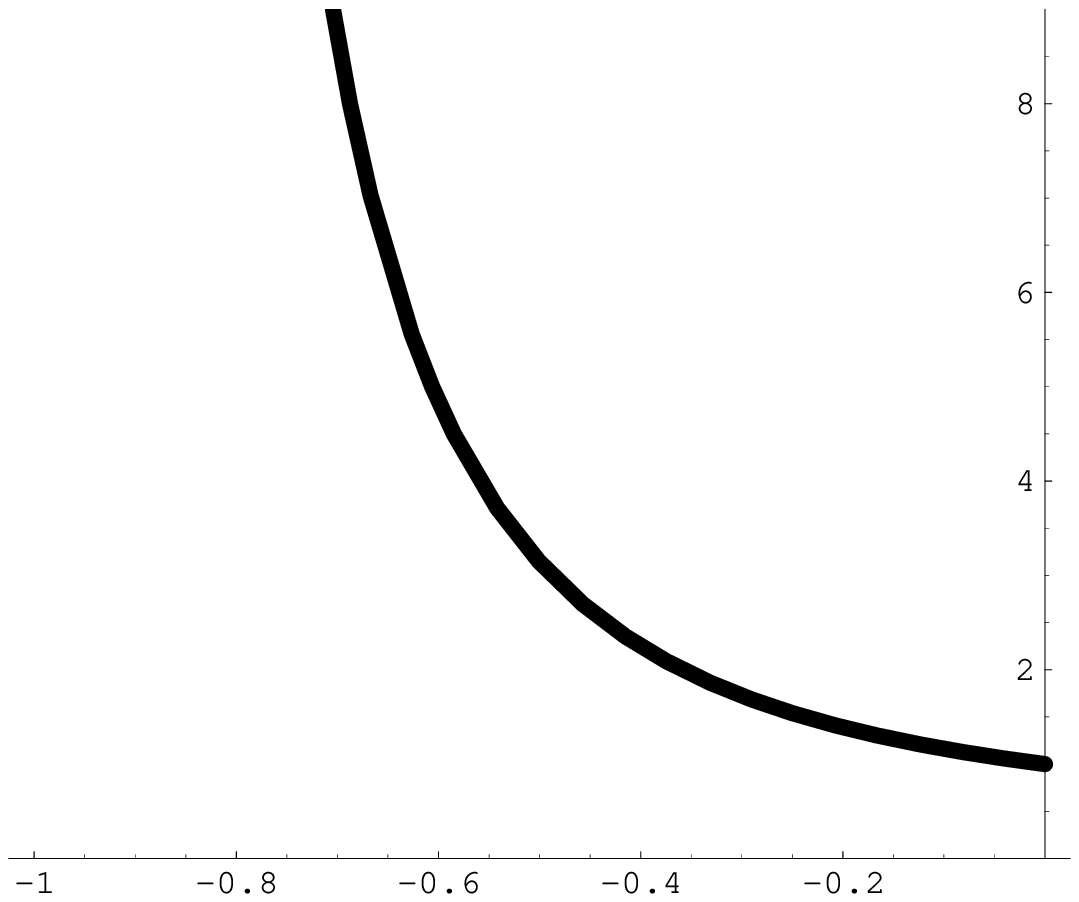}}
\put(134,100){\makebox(0,0){$D(z)$}}
\put(130,6){\makebox(0,0){$z$}}
\end{picture}
\caption{}
\label{D.eps}
}
\end{center}
\end{figure}

Next, we compute the Hamiltonian as in \cite{Se3}.
In this paper, we focus on the time dependence of
the tachyon field, and mainly assume $\del_iT=0$
for $i=1,\dots,p$.
Then the momentum conjugate to $T$ is
\beq
\Pi=\frac{\delta S}{\delta\dot T}
= \tilde{T}_p\, e^{-\frac{T^2}{4}} \alpha' \dot T {\mathcal F}'(z),
\eeq
where $\dot T$ stands for time derivative of $T$
and $z =-\frac{\alpha'}{2}\dot T^2$.
The Hamiltonian density is
\beq
{\cal H}=T_{00}= \tilde{T}_p\,  e^{-\frac{T^2}{4}}
D(z),
\label{Hami}
\eeq
where
\beq
D(z)={\mathcal F}(z)-2 z {\mathcal F}'(z) = -{\mathcal F}(z) \left(
1+2 \log 4\, z +4 z (\psi(z)-\psi(2z))
\right).
\eeq
where $\psi(z)$ is the poly-gamma function defined as
\beq
\psi(z) \equiv \frac{d}{dz} \log \Gamma(z)
= \frac{\Gamma'(z)}{\Gamma(z)}.
\eeq
The behavior of the
functions $D(z)$ and $\cF(z)$ for $-1<z\le 0$
is depicted in Figure \ref{F.eps} and Figure \ref{D.eps}

We would like to consider a solution of the equations of motion
which represent the tachyon rolling down to the bottom of
the potential. So, we require
$T^2 \rightarrow \infty$ at $x^0 \sim \infty$.
Then, (\ref{Hami}) implies that
$D(z)$ should become infinity in the limit $x^0\ra\infty$,
since the Hamiltonian $H= \int dx^{p} {\cal H}$ is conserved.
Therefore, it is important to examine the singularity
of $D(z)$ to know the asymptotic behavior of the solution.

Now, we summarize some properties of $\Gamma(z)$.
It is holomorphic except $z=-n$ with $n=0,1,\ldots$,
and has a pole
$\Gamma(z) \sim \frac{(-1)^n}{n !} \frac{1}{z+n}$
near $z=-n$.
And $1/\Gamma(z)$ is holomorphic in $|z| < \infty$.
The poly-gamma function $\psi(z)$
has singularities
$\psi(z) \sim \frac{-1}{z+n}$ at $z=-n$
with $n=0,1,\ldots$.
Using these properties, it is easy to see
that $\cF(z)$ and $D(z)$ behaves as
\begin{eqnarray}
\cF(z) &\sim& -n 4^{-n} \frac{(2n)!}{(n!)^2} \frac{1}{z+n},
\label{F}\\
D(z) &\sim& \frac{2n^2}{4^n}\frac{(2n)!}{(n!)^2}
\frac{1}{(z+n)^2},
\label{D}
\end{eqnarray}
near $z=-n$ where $n=1,2,\ldots
$\footnote{
Note that $z=0$ is a regular point for $\cF(z)$
and $D(z)$, as we have already seen in (\ref{asF}).}.

The fact that $D(z)$ has singularities for finite $z$
is crucial for our consideration.
Indeed, if $D(z)$ had no singularities except $z=\infty$,
the tachyon might become infinite within finite time.
Actually, we can show that it does occur for
the effective action of the Minahan-Zwiebach model \cite{MiZw} which has
${\mathcal F}(z)= 1+2 \log2\, z$, i.e.
\beq
S = -\tilde T_p\int d^{p+1} x e^{-\frac{T^2}{4}} \left(
1+ \log2 \, \alpha'\pa_\mu T \pa^\mu T \right),
\eeq
which is obtained by dropping
${\cal O}((\pa_\mu T \pa^\mu T)^2)$ terms
in (\ref{kinpo}).
In this case, $H=\tilde{T}_p\, e^{-\frac{T^2}{4}}
(1+\alpha' \log2\, \dot{T}^2)$,
which means $\dot{T} \sim
\sqrt{\frac{H}{\tilde{T}_p} e^\frac{T^2}{4}-1}$.
Then we can see that
$t= \int^{T} dT' (\frac{H}{\tilde{T}_p} e^\frac{{T'}^2}{4}-1)^{-\frac{1}{2}}
\sim \int^{T} dT' \sqrt{\frac{\tilde T_p}{H}} e^{-\frac{{T'}^2}{8}} $
and then $t$ is finite at $T \rightarrow \infty$.

Let us consider the behavior of $D(z)$ near $z=-1$,
which is the nearest singularity from the origin $z=0$.
{}From (\ref{D}), we know that the function $D(z)$ behaves as
\beq
D(z) \sim \frac{1}{(z+1)^2},
\eeq
near $z=-1$.
As explained above, if we set the initial condition
for $z$ to be in $-1<z\le 0$,
we expect the asymptotic behavior of
the solution is $T \sim \pm x^0$
in the limit $x^0 \rightarrow \infty$ and $-\infty$
(we set $\alpha'=2$ for notational simplicity).

Then, suppose that
the asymptotic form of the solution at $x^0 \sim \infty$ is
of the form
\beq
T= a + x^0+ \epsilon(x^0),
\label{asysol}
\eeq
where $\epsilon(x^0)$ is considered to be a small perturbation.
The energy conservation condition implies
\begin{eqnarray}
\dot\epsilon(x^0)\propto e^{-\frac{(x^0)^2}{8}},
\end{eqnarray}
where $\dot\epsilon$ denotes the time derivative of $\epsilon$.
Then,
the asymptotic behavior of the pressure can be computed as
\beq
P= {\cal L}&=& -\tilde{T}_p\,  e^{-\frac{T^2}{4}}\cF(z)
\label{pressure}\\
&\propto& e^{-\frac{(x^0)^2}{8}}.
\eeq
Therefore, this analysis also suggest that
the system behaves as a pressureless gas
for large time, which is consistent with
the result in \cite{Se2}.
So, we conclude that the asymptotic solution (\ref{asysol})
represents the tachyon matter in BSFT.

Note, however, that there is a
subtle difference between the two analyses.
The pressure near the end of the rolling
given in  \cite{Se2}
using the boundary state
is given by
\beq
P \sim -e^{ - \alpha x^0}
\eeq
which can also be reproduced by using
the effective action of \cite{Ga,Be,Kl}
if we take a specific form of the tachyon potential \cite{Se3}.
The sign of the pressure near the end of the evolution
in our model is also different from Sen's one.
Our analysis suggests that the pressure approaches
zero from the positive side, while it is negative in \cite{Se2}.
There are several possible reasons for the discrepancy.
First, the boundary state of \cite{Se2} is different from
the solution considered here.
Then, the asymptotic form of pressure
might be depend on the
specific forms of the solutions.
Second, the massive modes ignored in the analysis
might change the asymptotic form since
the difference between the solution and
the linear profile could be affected by them.
There is another possibility that
the higher derivative term could change the asymptotic form.

In the above consideration, we only considered $z$ which
satisfies $-1<z\le 0$.
We can also consider an asymptotic solution such that $z$ approaches
$-1$ from the $z<-1$ side.
\footnote{
We thank J. Minahan for pointing out an error
in the argument about this solution in the first version of the paper.
}
This solution has a peculiar property.
Since the function $D(z)$ increases when $z$ approaches
$-1$ from the $z<-1$ side,
the tachyon slows down while it rolls down the potential.
The situation
is the same for the other singularities $z=-n$ ($n=1,2,\dots$) of the
function $D(z)$, and there are asymptotic solutions approaching
$z=-n$. In any cases the pressure converges to zero,
and the system end up with the tachyon matter at late time.

In principle,
we can construct a full solution connecting
its asymptotic behavior $T\sim \pm x^0$.
There are basically two types of solutions
distinguished by the sign of the asymptotic values of $T$,
namely, time-like kink solutions, which approach
$T \ra x^0$ (or $T\ra -x^0$) as $x^0\ra\pm\infty$.
and bounce solutions $T \ra |x^0|$
(or $T\ra -|x^0|$) as $x^0\ra\pm\infty$.
This can be seen from the equation of motion,
\begin{eqnarray}
\half T D(z)+2\ddot T D'(z)=0.
\label{eom1}
\end{eqnarray}
Note that the left hand side is proportional to $\dot \cH\dot T^{-1}$,
and hence it is equivalent to the energy conservation condition
as long as $\dot T\ne 0$.
Since we know $D(z)>0$ and $D'(z)<0$ from Figure \ref{D.eps},
(\ref{eom1}) implies that $T$ is accelerated toward the bottom
of the the potential. Thus, once the tachyon starts rolling down
the potential, it will never stop. Then, suppose that we set the initial
condition, such that the tachyon rolls up the potential.
Since $D(z)\ge 1$ and the equality holds only for $z=0$,
the energy conservation condition implies that
the tachyon cannot stop if
 the energy density is bigger
than the tension $\tilde T_p$ and get over the potential
barrier, while it will stop at the point where
$\tilde T_p \, e^{-\frac{T^2}{4}}$ is equal to the energy
and turn back if the energy density is smaller
than the tension $\tilde T_p$.

Note that both of these two types of solutions
are localized in the time direction.
The time-like kink solutions has ``charge"
\footnote{
This ``charge'' is not the usual conserved charge.
It is defined as the integral of the RR-field over a large
sphere surrounding the finite time region of the
non-BPS D-brane in the ten dimensional space-time.\cite{GuSt}
},
which is calculated from the
Chern-Simons term for the non-BPS D-brane \cite{KrLa,TaTeUe},
$S_{CS}= \tilde{T}_p \int{\rm Str}\, e^{2\pi \alpha' {\cal F} } \wedge C$,
where ${\cal F}$ is the superconnection.
Then, we can identify this type of solutions
as the space-like branes considered in \cite{GuSt}.

Note that this time-like kink solution cannot be obtained
by simple Wick rotation of usual kink solution for the BSFT
action.
The kink solution is of the form
$T=ux$ where $u$ is a parameter and we have to
take the limit $u\ra\infty$ to minimize the action.
The Wick rotation trick $u\ra iu$ and $x\ra ix^0$
implies a configuration with $z\sim -u^2 \ra -\infty$.
However, there is an upper bound
for the speed of the rolling of the tachyon
in the time-like kink solutions,
which is not compatible with this limit.


\section{Cosmological Evolution using the BSFT action}
\label{Cosmo}

Here, we sketch the cosmological evolution
by the BSFT action (\ref{bsfta}) following \cite{Gi}.
Although
it is difficult to make a precise analysis
in the string field theory point of view,
since the higher derivative terms are truncated
in the action, it is still interesting to see
the qualitative nature of the model
and compare it with other models
for the cosmological evolution driven by
the rolling tachyon \cite{Gi}.
(See also \cite{Mu}--\cite{ShWa}.)

We consider the BSFT action (\ref{bsfta})
in 4 dimension,
turning on the minimal coupling to gravity,
with the usual Einstein-Hilbert action.
We assume
that the tachyon only depends on time
and the metric is the Robertson-Walker metric
\begin{eqnarray}
ds^2=-dt^2+a(t)\left(
\frac{dr^2}{1-Kr^2}+r^2 d\Omega_2^2
\right).
\end{eqnarray}
The equations of motion are
\begin{eqnarray}
\left(\frac{\dot a}{a}\right)^2&=&
\frac{8\pi G}{3}\rho-\frac{K}{a^2},
\label{E}\\
\frac{\ddot a}{a}&=&-\frac{4\pi G}{3}(\rho+3P),
\label{acc}
\end{eqnarray}
where $\rho$ and $P$ are the energy density and
pressure given in (\ref{Hami}) and (\ref{pressure}), respectively.
\begin{eqnarray}
\rho&=& \tilde{T}_p\,  e^{-\frac{T^2}{4}}D(z),
\label{rho}\\
P&=& -\tilde{T}_p  e^{-\frac{T^2}{4}}\cF(z),
\label{P}
\end{eqnarray}
where $z=-\dot T^2$.

{}From (\ref{E}),
we can immediately see that $\dot a$ is always positive
for $K\le 0$ cases,
describing the expanding universe,
since the function $D(z)$ is positive
the weak energy condition $\rho>0$ is satisfied.

The equations of motion also imply
\begin{eqnarray}
\dot\rho=-3\frac{\dot a}{a}(\rho+P),
\label{eqmr}
\end{eqnarray}
which is equivalent to the
conservation equation for the energy momentum tensor.
Inserting (\ref{rho}) and (\ref{P}),
this equation can be rewritten as
\begin{eqnarray}
\half T D(z)+2\ddot T D'(z)=3\frac{\dot a}{a}
(D(z)-\cF(z))\dot T^{-1}.
\label{eom}
\end{eqnarray}
Suppose that we set the initial condition $T(\bar{t})>0$ and
$\dot T(\bar{t})=0$ at $x^0=\bar{t}$.
We can easily see that the right hand side
converges to zero in the limit $\dot T\ra 0$.
Then, $D(0)=1$ and $D'(0)= - F'(0) \sim -2\log2<0$
imply $\ddot T>0$ which
shows that the tachyon starts rolling down the potential.
This argument also implies that the tachyon will not stop rolling
once it starts, as we observed in the previous section for
the model without gravity.

\begin{figure}[htb]
\begin{center}
\unitlength=.4mm
\begin{picture}(150,100)(0,0)
\epsfxsize=5cm
\put(0,0){\epsfbox{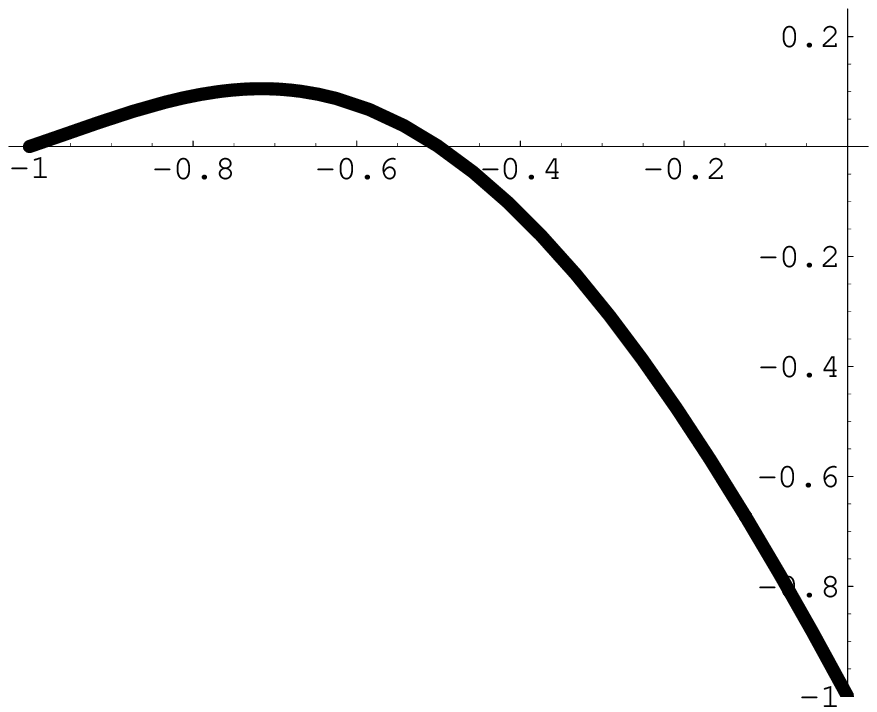}}
\put(136,100){\makebox(0,0){$\omega(z)$}}
\put(130,82){\makebox(0,0){$z$}}
\end{picture}
\end{center}
\caption{}
\label{O.eps}
\end{figure}

The ratio of the pressure and the energy density is given by
\beq
\omega=\frac{P}{\rho}=\frac{1}{1+2 \log4\, z + 4z (\psi(z)-\psi(2z))},
\eeq
which has maximum near $z \sim -0.7$ with $\omega(-0.7) \sim 0.1$
and $\omega(0)=-1, \omega(-1)=0$ as depicted in Figure \ref{O.eps}.
Hence, the dominant energy condition $|\omega| \leq 1$
is satisfied. However
the strong energy condition
\beq
\rho+3P \geq 0,
\eeq
does not hold for $|z|\leq 0.278\cdots$.
As we can see from the equation (\ref{acc}),
this implies  the acceleration of the scale factor
in the initial stage with small $|z|$.

In summary,
the qualitative behavior of the model
is roughly the same as that given in \cite{Gi}.
It would be interesting to examine
further to see if it can be
considered as a realistic model of inflation
or dark matter. (See \cite{Mu}--\cite{ShWa})

\section{Conclusion and Discussion}
\label{Conc}

In this paper,
we have analysed the classical decay process of
unstable D-branes in superstring theory using
the BSFT action.
We have shown that the solutions of the equations of motion
for the tachyon field asymptotically approach to $T=\pm x^0$
producing the tachyon matter.
We also considered the cosmological evolution
driven by the rolling tachyon
using the BSFT action.

We used the BSFT action obtained by dropping
terms which include second or higher derivatives of
the tachyon field. However, 
in the time dependent solution of the equations of motion for
this effective action, the second or higher derivatives of
the tachyon field is not precisely zero. Therefore,
there could be some corrections for the analysis
if we include the terms we dropped.

Let us make some comments about the possible corrections.
First, recall that the BSFT action that we used can be trusted
for the tachyon of the form $T=a+u x^0$.
We showed that the tachyon approaches the linear profile
$T\sim\pm x^0$ in the asymptotic region.
This comes from the fact that the function $D(z)$ has a
pole at $z=-1$. The function $D(z)$ could receive some
corrections. However, we expect that even if we
include these corrections, it
still has a singularity for the tachyon
configuration $T=a \pm x^0$.
This is because the action (\ref{bsfta}) is exact for
the linear profile and we know that the action has a singularity
for this tachyon configuration.
It is quite natural to expect that the singularity
coincides for the action and the energy density.
Therefore, we strongly believe that
there are solutions which approach
$T\sim\pm x^0$ in the limit $|x^0|\ra\infty$
even if we include the corrections.

On the other hand, we cannot
exclude the possibility that
there are solutions which does not
approach $T\sim\pm x^0$
when we include the higher derivative corrections.

The precise form of the solution for the tachyon field
seems to be difficult to obtain in this approach.
In addition to the complexity of the action,
there are some more fundamental difficulties.
If we include the higher derivative corrections,
the massive modes can no longer be ignored \cite{KuMaMo1}.
The BSFT action including these modes has not been
obtained because it corresponds to the
non-renormalizable terms in the boundary interaction.
Therefore, our solution cannot be trusted away from
the asymptotic region.
However, most of our qualitative arguments given in
section \ref{Tmatter} and section \ref{Cosmo}
only require the broad behavior of the functions
$\cF(z)$ and $D(z)$ and so we hope that
these analyses successfully catch the essential features
of the theory.

We have observed that the asymptotic behavior of the pressure
is slightly different from that obtained in \cite{Se2}.
These ambiguities might be the origin of this discrepancy.
In order to clarify these issues,
it would be important to study the relations between
the BSFT approach and
the boundary state approach given in \cite{Se1,Se2}.

In this paper we only turned on the tachyon field.
The action including both the gauge field strength $F$ and the tachyon
for a non-BPS D-brane was obtained in a closed form \cite{TeUe} as
\beqa
\label{BSFTaction}
S&=&
-\sqrt{2}T_9\int d^{10}x e^{-\frac{1}{4}T^2}\sqrt{-\det(\eta_{\mu\nu}
+F_{\mu\nu})}{\mathcal F}\left(\frac{1}{4\pi}{\mathcal
G}^{\mu\nu}\pa_{\mu}T\pa_{\nu}
T\right),
\eeqa
where
\beqa
{\mathcal G}^{\mu\nu}\equiv\left(\frac{1}{1+F}\right)^{(\mu\nu)}.
\eeqa
We can also obtain the action including scalar fields, representing
the transverse fluctuation of the D-brane, by T-dualizing
the above action.
It would be interesting to include
the effect of
the gauge field and scalar fields in the consideration
of the rolling tachyon. See \cite{Ha} for the related analysis.
In particular, the time evolution of the system
with the electric flux \cite{GiHoYi} is interesting to study.

As argued in \cite{Te,AsSuTe}, 
all the D$p$-branes can be constructed
from infinitely many unstable D-particles,
i.e. non-BPS D0-branes in type IIB theory
or D0-brane - anti D0-brane pairs in type IIA theory.
Therefore, it is possible to analyse the
decay process of the non-BPS D$p$-brane
in terms of the unstable D-particles.
Then, the tachyon matter produced by the decay of
the non-BPS D$p$-brane can be thought of as
the pressureless gas consists of the decay
products of the unstable D-particles.
It might be interesting to consider the
tachyon matter in this point of view.

\vskip6mm\noindent

\subsection*{Note added}

After this work was completed,
a related paper \cite{Mi} appeared.

\subsection*{Acknowledgments}

\vskip2mm
We would like to dedicate this article 
to the memory of Sung-Kil Yang
from whom we learned a lot of things.
We will not forget his warm encouragements.

S.T. would like to thank
J. de Boer for helpful discussions and
T. Takayanagi and J. Minahan
for useful comments and discussions.
S.S. is grateful to T. Okuda and K. Okuyama for
useful discussions.

\end{document}